# Gender differences in research collaboration[1]

Giovanni Abramo

*National Research Council of Italy (IASI-CNR) and Laboratory for Studies of Research and Technology Transfer at University of Rome "Tor Vergata" – Italy*

ADDRESS: Dipartimento di Ingegneria dell'Impresa, Università degli Studi di Roma "Tor Vergata", Via del Politecnico 1, 00133 Roma – ITALY
tel. and fax +39 06 72597362, giovanni.abramo@uniroma2.it

Ciriaco Andrea D'Angelo

*Laboratory for Studies of Research and Technology Transfer at University of Rome "Tor Vergata" – Italy*

ADDRESS: Dipartimento di Ingegneria dell'Impresa, Università degli Studi di Roma "Tor Vergata", Via del Politecnico 1, 00133 Roma – ITALY
tel. and fax +39 06 72597362, dangelo@dii.uniroma2.it

Gianluca Murgia

*University of Siena – Italy*

ADDRESS: Dipartimento di Ingegneria dell'Informazione e Science Matematiche, Università degli Studi di Siena, Via Roma 56, 53100 Siena – ITALY
tel. and fax +39 0577 1916386, murgia@dii.unisi.it

# Gender differences in research collaboration

**Abstract**

The debate on the role of women in the academic world has focused on various phenomena that could be at the root of the gender gap seen in many nations. However, in spite of the ever more collaborative character of scientific research, the issue of gender aspects in research collaborations has been treated in a marginal manner. In this article we apply an innovative bibliometric approach based on the propensity for collaboration by individual academics, which permits measurement of gender differences in the propensity to collaborate by fields, disciplines and forms of collaboration: intramural, extramural domestic and international. The analysis of the scientific production of Italian academics shows that women researchers register a greater capacity to collaborate in all the forms analyzed, with the exception of international collaboration, where there is still a gap in comparison to male colleagues.

**Keywords**

*Research collaboration; collaboration patterns; gender; bibliometrics; university.*

# 1. Introduction

The scientific debate on gender aspects in research systems has focused primarily on the overrepresentation of male academics, the productivity gap and issues of gender discrimination. Less attention has been given to gender differences in research collaboration, in spite of the fact that scientific research is increasingly conducted in collaboration (Archibugi and Coco 2004; Katz and Martin 1997) and that positive correlation has been demonstrated between intensity of collaboration and scientific productivity (Abramo, D'Angelo, and Di Costa 2009). The major studies in this area have been conducted through surveys that have primarily attempted to identify the causes of the different collaboration behaviors of male and female researchers (Bozeman and Gaughan 2011; Rhoten and Pfirman 2007). However studies that have analyzed the collaborations initiated by male and female researchers by beginning from the resulting scientific production are rare and focused on limited contexts (Badar, Hite, and Badir 2013; Long 1992).

In this work we propose to advance knowledge on the theme through an innovative approach based on large-scale bibliometric analysis at the individual level. In particular, we intend to verify if gender differences actually exist in the propensity to collaborate, also searching for potential differences in the forms of collaboration (intramural, extramural: domestic/international) and across disciplines and fields. The material for the analysis is the publications indexed in Web of Science (WoS). While these do not represent the only output of scientific collaborations, co-authored publications do provide a meaningful proxy, and can be easily measured without any bias related to the object of analysis or the actor who carries out the data collection (Melin and Persson 1996). With reference to the Italian higher education system, we are able to attribute each publication to the academic that produced it, with a very low error rate. Through the scientists' institutional affiliations, we are able to discriminate intramural from extramural collaborations, and among the latter, domestic from international. To verify to what extent any gender differences in propensity to collaborate depend on the authors' specializations in a particular field or discipline, we take advantage of a unique characteristic of the Italian university system, where each academic is classified into one and only one Scientific Disciplinary Sector (SDS). There are 370 such fields[2], in turn grouped under 14 University Disciplinary Areas (UDAs).

The empirical evidence revealed by the study provides support for the policy-makers responsible for national research systems and individual universities, in designing and evaluating the effectiveness of the policies intended to favor gender equity in collaboration (Rees 2002; Zippel 2011).

After reviewing the relevant literature in Section 2, in the next section we describe the methodology applied and the field of observation. In Section 4 we show the results obtained and in Section 5 we indicate questions for further examination and provide several policy indications.

---

[2] The complete list is accessible at http://attiministeriali.miur.it/UserFiles/115.htm, last accessed on March 15, 2013.

## 2. Gender aspects in research collaboration: literature review

A number of studies in the literature, generally dating to some years ago, have reported a tendency among female academics to collaborate in a different and less effective manner compared to their male colleagues (Cole and Zuckerman 1984; Sonnert 1995). Indications are that female academics tend to develop more formal collaborations (Sonnert 1995) and networks of contacts that are less "cosmopolitan" (Bozeman and Corley 2004) and less prestigious (Fuchs, von Stebut, and Allmendinger 2001; Long 1990). Furthermore, women tend to broaden their individual networks beyond their own specialties (Leahey 2006). This can limit their level of specialization, although it can also favor greater propensity to interdisciplinary collaboration (Rhoten and Pfirman 2007; van Rijnsoever, Hessels, and Vandeberg 2008). Even if studies on gender differences in collaboration behavior show conflicting results (Bozeman and Gaughan 2011; Bozeman and Corley 2004; Cole and Zuckerman 1984; Fox 1991), the literature mainstream reports a gap in the social capital of women academics (Rhoten and Pfirman 2007), which can be further aggravated by the so-called "Matilda effect" (Rossiter 1993). The Matilda effect, a counterpart of the famous "Matthew effect in science" (Merton 1968), occurs when women researchers, in spite of equal or greater contribution to a paper than male colleagues, are not recognized in the resulting publication bylines.

On the other hand, collaborations have a very strong impact on the productivity of women researchers in comparison to the impact for men (Badar, Hite, and Badir 2013; Kyvik and Teigen 1996). In fact, collaborations permit women to overcome the lack of social capital and better integrate in the academic environment. It is no accident that in the area of the economic disciplines, women academics tend to collaborate primarily in the first years of their career (McDowell and Smith 1992; McDowell et al. 2006). In contrast, male colleagues place more emphasis on building a solid reputation through the independent production of publications, in order to later achieve better collaborations. Thus the choices made by women can cause a slowdown in the process of developing their scientific reputations and lesser capacity to publish works in more prestigious journals (Breuning and Sander, 2007; Grant and Ward 1991).

Female academics' lack of social capital is caused first of all by mechanisms of gender homophily that stimulate a search for collaborations primarily among colleagues of the same gender, with whom one is more likely to share values and methodological approaches (Boschini and Sjogren 2007; Ferber and Teiman 1980; McDowell and Smith 1992). Because of the fact that women academics are still a minority in the principle disciplines, these mechanisms lead to greater isolation compared to the situation for men (Rivellini, Rizzi, and Zaccarin 2006).

The isolation of women is made still more acute by the fact that in the academic environment, because of the historic overrepresentation of men, an essentially masculine culture and value system frequently dominate (Rhoten and Pfirman 2007). Isolation of women academics is greater in smaller departments (McDowell and Smith 1992) and in those with lower percentages of women (Etzkowitz, Kemelgor, and Uzzi 2000). Isolation is also more notable in disciplines dominated by "old boy networks", where scientific debate, reputations of scientists, access to collaboration and research funds are heavily influenced by informal networks composed exclusively of male academics (Fox 1991; O'Leary and Mitchell 1990). Especially in these contexts, women receive less social support, less professional recognition, exercise less power and find less opportunity to collaborate compared to their male colleagues (Etzkowitz, Kemelgor, and Uzzi 2000). This explains the tendency of women to concentrate, to significantly measurable extent, in academic activities that are less linked to research, such as teaching and administration (Olsen, Maple, and Stage 1995; Taylor, Fender, and Burke 2006).

The isolation of women academics already develops in the phase of entry to the academic environment, so much so that they generally evaluate their mentors as less satisfactory than do their male colleagues (Sambunjak, Straus, and Marusic 2006). Their perception could be partly due to the greater expectations that women hold for their mentors' actions (Seagram, Gould, and Pyke 1998). In other cases, isolation is due to the discrimination to which women academics may fall victim, particularly in regards to their family choices. For example, the decision not to marry reduces possibilities of collaboration with a mentor – who could be compromised by associating with an "available" colleague – or the choice to have children, which could be interpreted as a lack of interest in one's academic career (Long 1990). Beyond cases of discrimination, women do not feel adequately involved by their mentors in preparation of scientific articles (Seagram, Gould, and Pyke 1998) or in socialization processes essential for academic success (O'Leary and Mitchell 1990). Further, women are often entrusted to mentors outside their home institution (Fuchs, von Stebut, and Allmendinger 2001) or go completely without one (Ledin et al. 2007). Studies on the effective levels of collaboration developed between women academics and mentors have given contrasting results (Borrego et al. 2008; Feldman 1974), although mentorship clearly impacts on productivity of women (Long 1990) and on their possibilities to access research funds (NRC 2009). For this reason, various institutions have developed mentorship programs directed at gender equity (Mark et al. 2001) and women themselves tend to prefer collaboration with professors who have implemented equality strategies (Bozeman and Corley 2004).

The success of these programs, together with the growth in the percentage of women in the academic environment, could help reduce their shortfall in social capital and fully exploit their potentially greater propensity for group work, which has been noted in some organizational psychology studies (Keashly 1994).

Until now, studies on gender differences in the propensity to collaborate show conflicting results (Abramo, D'Angelo, and Caprasecca 2008; Moya Anegón et al. 2009; McDowell and Smith 1992; McDowell et al. 2006). Focusing on the specific forms of collaboration, women seem to have more propensity to collaborate with colleagues belonging to other domestic institutions (Moya Anegón et al. 2009). Differently, male colleagues register a greater propensity to collaborate at the international level (Frehill, Vlaicu, and Zippel 2010; Larivière et al. 2011), even if Melkers and Kiopa (2010) show that this gender difference, among U.S. academics involved in stem cell research, is not statistically significant The lower propensity to collaborate at the international level by female researchers could be due to the lack of research funds, which limits the breadth of collaboration networks for the individual academic (Bozeman and Corley 2004), since it reduces the possibility of paying the costs of transfers and thus the capacity to attract international fellows (Liang et al. 2009). The lack of funding availability for female researchers can be attributed in part to the existence of gender bias in the peer-review procedures adopted for the evaluation of project proposals (Bornmann, Mutz, and Daniel 2007; Ledin et al. 2007; Wenneras and Wold 1997). Another obstacle to international collaboration for women is presented by the prejudices against women that still exist in certain countries and which lead local researchers, who are primarily men, to avoid undertaking collaborations with women colleagues (Frehill, Vlaicu, and Zippel 2010). Finally, international collaborations initiated by women academics are strongly conditioned by their family ties, which limit collaboration duration and geographic extension (Frehill, Vlaicu, and Zippel 2010). Such restrictions seem to influence male academics less, who continue to travel abroad with advancing age, more than women do (Zimmer, Krimmer, and Stallmann 2006). In the case of women academics, the presence of children (Shauman and Xie 1996) and a working husband (Ackers 2004) represent the principle restrictions on international mobility. In particular, the absence of adequate policies for "partner mobility" limits female researchers much more strongly than it does their male colleagues, who tend to be less involved in dual-career families (Zippel 2011) or are in such an advanced phase of career that they can comfortably arrange for their spouse to follow (Rosenfel and Jones 1987). The impact of reduced mobility due to family has a much stronger impact on the careers of women academics. In fact, since women tend not to be inserted in old-boy networks, they have fewer instruments, such as conferences, to initiate collaborations with internationally-recognized colleagues (Poole, Bornholt, and Summers 1997). It is no accident that women academics perceive family obligations as a strong restriction on international collaborations (Fox 2009), and thus tend to marry less frequently than do their male colleagues (Marwell, Rosenfeld, and Spilerman 1979; Shauman and Xie 1996).

These limitations might be overcome thanks to the spread of new communications technologies, which permit initiations of international collaborations at lower cost and with less impact on family life. Still, various authors remain skeptical about the possibilities for women to develop truly effective international collaborations through the use of these technologies (Campion and Shrum 2004) and the few studies conducted thus far have given conflicting results (Ding et al. 2010; Miller et al. 2006). The analysis of relationships between gender and form/intensity of collaboration must also take account of certain variables in context, such as the field of specialization, which can have characteristics that strongly affect both men's and women's propensities to collaborate (Bozeman and Gaughan 2011; Moya Anegón et al. 2009).

The current study differs from the preceding ones in several aspects: i) beyond detecting possible gender differences in the propensity to collaborate in general and at domestic and international levels, it also investigates intramural collaborations; ii) the field of analysis is not represented by a sample, rather by the entire population of Italian academics in the hard sciences and economics (43,371 observations), which makes inferential analysis less important; iii) gender differences are investigated at discipline and field levels (11 fields subdivided in 200 levels); and least but probably most important iv) the methodology adopted (see next section), using the single scientist as the base unit of analysis, avoids the inevitable distortions embedded in bibliometric analyses at the aggregate level.

## 3. Methodology, dataset and indicators

*3.1 Methodological approach*

The analysis of gender differences in research collaboration has primarily been conducted by surveys (Bozeman and Gaughan 2011; Rhoten and Pfirman 2007), while the use of bibliometric data has generally been limited to statistical validations of hypotheses, such as concerning gender homophily (McDowell and Smith 1992). The very few studies (Lariviere et al. 2011; Moya Anegón et al. 2009) conducted with bibliometric data have used a methodology based on the aggregation of publications by gender of author, meaning that the propensity to collaborate for either gender is calculated by dividing the total number of co-authored publications by the gender by the total number of their publications. Since the distribution of scientific production is in general very skewed[3], calculating the propensity to collaborate in this manner will very likely give values affected by the presence of outliers and thus not reflect the true propensity of the large part of researchers. As shown by Abramo, D'Angelo and Caprasecca (2009), gender differences in research performance are notable only among top 10% scientists and negligible for all the others, which confirm the would-be distortions of aggregate level analyses.

An alternative approach for study of co-authorship, first proposed by the current authors (Abramo, D'Angelo, and Murgia 2013), consists of using the single scientist as the base unit of analysis. It entails calculating the propensity to collaborate of each scientist, as the ratio of his or her co-authored publications to the total of his or her publications. The female propensity to collaborate will then be the average of the individual female propensities. Differently from aggregate level analyses, our approach is therefore much more accurate in defining the gender traits, as it limits the distortions due to the presence of outliers. As shown by Abramo, D'Angelo, and Murgia (2013), the values of collaboration propensities for the scientists could be quite different from those obtained under the previously-used approach. Our analysis refers to all Italian university professors in the hard sciences and some fields of the social sciences, where publications indexed by bibliometric databases represent a good proxy of overall research output (Moed 2005). We exclude the arts and humanities, where the coverage of bibliometric databases is too limited. The instrument of analysis is the co-authorship of scientific publications over the period 2006-2010 as indexed on the Web of Science (WoS). For each form of collaboration, we first calculate the different propensities to collaborate for the individual female and male scientists, then analyze differences across disciplines, and finally across fields within each discipline. Our method also allows us to test the correlation between the propensities to collaborate in the various forms.

*3.2 Data sources and field of observation*

The dataset of Italian professors used in our analysis has been extracted from a database[4] maintained by the Ministry of Education, Universities and Research (MIUR). This database indexes the names, academic rank, discipline (SDS/UDA), and institutional affiliation of all academics in Italian universities. We identify gender by examining the academics' first names and in cases of doubt verify by checking personal and institutional Web pages.

---

[3] In the Italian case 23 percent of academics produce 77 percent of overall scientific advancement, Abramo, Cicero, and D'Angelo 2013.

[4] http://cercauniversita.cineca.it/php5/docenti/cerca.php, last accessed on March 15, 2013.

Next, the dataset of these individuals' publications is extracted from the Italian Observatory of Public Research (ORP), a database developed and maintained by the authors and derived under license from the WoS. Beginning from the raw data of 2006-2010 Italian publications in the WoS, and applying a complex algorithm for disambiguation of the true identity of the authors and their institutional affiliations (for details see D'Angelo, Giuffrida, and Abramo 2011), each publication[5] is attributed to the university scientist or scientists (full, associate and assistant professors) that produced it, with a harmonic average of precision and recall (F-measure) equal to 96 (error of 4 percent).

For each publication, the bibliometric dataset thus provides:
- the complete list of all coauthors;
- the complete list of all their addresses;
- a sub-list of only the academic authors, with their gender, SDS/UDA and university affiliations.

Our dataset permits unequivocal identification of each academic with their home university, although this operation is not possible for non-academic authors of the publications. It is also not possible to associate the academics with any organizations other than their own universities, although the literature shows (Katz and Martin 1997) that in some cases authors indicate more than one institutional address, due to some form of multiple engagement or change in employment. This can actually lead to certain problems, such as classifying publications as being produced under international co-authorship when the presence of a foreign organization in the byline is actually due to a single academic belonging to multiple organizations (Glanzel 2001)[6]. Further, our dataset permits unequivocal assignment of every academic to their SDS, and thus to the UDA to which they belong, while the same operation is not possible for non-academic authors of the publications. For these reasons, the analysis is conducted only for university researchers.

Table 1 presents the statistics for the population of Italian academics belonging to the 11 UDAs analyzed, with their respective publications. To render the bibliometric analysis still more robust, the field of observation is limited to those SDSs (200 in all) where at least 50 percent of academics produce at least one publication in the 2006-2010 period.

The number and percentage of female academics in the composition of each UDA varies substantially, from 12.8 percent in Industrial and information engineering to 54.5 percent in Pedagogy and psychology. The comparison of these percentages with those for the years 2001 to 2003 years (Abramo, D'Angelo, and Caprasecca 2008) highlights an increase of the presence of women in all the UDAs, except for Mathematics and computer sciences, in which the percentage of women has decreased slightly. Biology is the only UDA where more than 50 percent of publications have at least one female author; however in this UDA the percentage of publications with at least one male author is a further 20 points higher. In substance, differing from previous studies conducted for Italy, France and

---

[5] We exclude those document types that cannot be strictly considered as true research products, such as editorial material, conference abstracts, replies to letters, etc.

[6] In Section 3.3 we describe the methodological assumptions that address this critical problem.

Spain (Moya Anegón et al. 2009; Naldi and Vannini Parenti 2002), our data reveal that the large majority of the dataset of publications (almost two thirds) do not indicate any women academics in the author byline. Also different is that in all the UDAs, with the exception of Pedagogy and psychology, Biology and Economics and statistics, there are less than 20 percent of publications without a male academic in the byline.

The percentage of "productive" academics (producing at least one publication indexed under the WoS in the period 2006-2010) differs between female and male academics. The productive female academics are 84.4 percent of the total, while the productive male academics are 83.1 percent of the total. The difference between the percentage of the productive female and male academics peaks at 6.2 percent in Industrial and information engineering, and falls to -6.0 percent in Mathematics and Computer Sciences.

The percentage of "collaborative" academics (at least one publication in co-authorship with other scientists in the same period) also differs between female and male academics. The collaborative female academics are 83.7 percent of the total, while the collaborative male academics are 82.0 percent of the total. The difference between the percentage of the collaborative female and male academics peaks at 6.5 percent in Industrial and information engineering, and falls to -5.5 percent in Mathematics and computer sciences.

*Table 1: Principle characteristics of the population of academics analyzed*

| UDA | Gender | Publications | Research staff | | |
| --- | --- | --- | --- | --- | --- |
| | | | Total | Productive | Collaborative |
| Mathematics and computer sciences (MAT) | F | 4,867 (30.2%) | 1,202 (33.3%) | 920 (76.5%) | 892 (74.2%) |
| | M | 13,259 (82.2%) | 2,405 (66.7%) | 1,985 (82.5%) | 1,917 (79.7%) |
| Physics (PHY) | F | 6,359 (26.8%) | 503 (17.5%) | 453 (90.1%) | 450 (89.5%) |
| | M | 21,880 (92.3%) | 2,370 (82.5%) | 2,149 (90.7%) | 2,125 (89.7%) |
| Chemistry (CHE) | F | 12,479 (48.6%) | 1,357 (37.6%) | 1,288 (94.9%) | 1,288 (94.9%) |
| | M | 21,804 (84.9%) | 2,253 (62.4%) | 2,096 (93.0%) | 2,079 (92.3%) |
| Earth sciences (EAR) | F | 1,526 (28.9%) | 349 (24.5%) | 294 (84.2%) | 292 (83.7%) |
| | M | 4,553 (86.2%) | 1,074 (75.5%) | 887 (82.6%) | 880 (81.9%) |
| Biology (BIO) | F | 16,468 (52.7%) | 2,862 (48.9%) | 2,567 (89.7%) | 2,562 (89.5%) |
| | M | 23,867 (76.3%) | 2,993 (51.1%) | 2,677 (89.4%) | 2,663 (89.0%) |
| Medicine (MED) | F | 21,662 (34.4%) | 3,394 (27.3%) | 2,804 (82.6%) | 2,796 (82.4%) |
| | M | 57,662 (91.5%) | 9,039 (72.7%) | 7,380 (81.6%) | 7,318 (81.0%) |
| Agricultural and veterinary sciences (AVS) | F | 5,822 (49.5%) | 1,018 (32.0%) | 911 (89.5%) | 910 (89.4%) |
| | M | 9,894 (84.1%) | 2,165 (68.0%) | 1,809 (83.6%) | 1,804 (83.3%) |
| Civil engineering (CEN) | F | 919 (17.1%) | 277 (15.9%) | 200 (72.2%) | 196 (70.8%) |
| | M | 4,960 (92.3%) | 1,470 (84.1%) | 1,030 (70.1%) | 1,012 (68.8%) |
| Industrial and information engineering (IIE) | F | 6,958 (18.7%) | 720 (12.8%) | 657 (91.3%) | 655 (91.0%) |
| | M | 34,765 (93.2%) | 4,924 (87.2%) | 4,189 (85.1%) | 4,161 (84.5%) |
| Pedagogy and psychology (PPS) | F | 1,614 (48.3%) | 575 (54.5%) | 385 (67.0%) | 377 (65.6%) |
| | M | 2,347 (70.2%) | 480 (45.5%) | 330 (68.8%) | 327 (68.1%) |
| Economics and statistics (ECS) | F | 1,189 (33.2%) | 631 (32.4%) | 397 (62.9%) | 369 (58.5%) |
| | M | 2,725 (76.1%) | 1,318 (67.6%) | 803 (60.9%) | 731 (55.5%) |
| Total | F | 73,785* (37.4%) | 12,888 (29.7%) | 10,876 (84.4%) | 10,787 (83.7%) |
| | M | 178,689* (90.5%) | 30,491 (70.3%) | 25,335 (83.1%) | 25,017 (82.0%) |

\* Totals are less than the sum of the column data due to double counts of publications co-authored by female (male) academics where the research subject pertains to more than one UDA.

*3.3 Indicators and methods*

Beginning from the individual academic of known gender and SDS, we will compare the average propensity to collaborate in the different fields for each of four forms: in general, intramural, and extramural with researchers from domestic and from foreign organizations. The first form of collaboration, the propensity to collaborate in general, represents a superset of the other forms.

We construct an "author-publication" matrix of dimensions *m* x *n*, with:
- m = 36,211, i.e. total number of productive academics,
- n = 197,460, i.e. total number of their publications.

We then associate each academic with his or her publications (*p*) over the period. Since for each publication we know the number of authors and the numbers of domestic and foreign institutions, for each scientist we can calculate the number of publications resulting from collaborations (*cp*), the number of publications resulting from collaborations with other academics belonging to the same university (intramural - *cip*), the number of publications from collaborations with scientists belonging to other domestic organizations (extramural domestic - *cedp*), and the number of publications with scientists belonging to foreign organizations (extramural international - *cefp*)[7]. From these values we can construct the indicators for the relative individual propensities to collaborate, from which we can then also obtain the average propensities per gender in each field and discipline:

- Propensity to collaborate $C = \frac{cp}{p}$
- Propensity to collaborate intramurally $CI = \frac{cip}{p}$
- Propensity to collaborate extramurally at the domestic level $CED = \frac{cedp}{p}$
- Propensity to collaborate extramurally at the international level $CEF = \frac{cefp}{p}$

Each of the four indicators varies between zero (if, in the observed period, the scientist under observation did not produce any publications resulting from the form of collaboration analyzed); and 1 (if the scientist produced all his or her publications through that form of collaboration).

## 4. Results and discussion

The calculation of *C*, *CI*, *CED* and *CEF* permits the comparison of the values of propensity registered for female and male academics, relative to different forms of co-authorship and in the different UDAs. The report on these steps of our analysis is presented in section 4.1. In section 4.2 we then examine gender differences in the correlations between pairs of the four indicators.

---

[7] Single-authored papers with more than one affiliation are not considered as collaborations. A publication with more than two authors could present different forms of collaboration, for example intramural and extramural domestic. In this case it is counted in calculating propensity for each form of collaboration observed.

*4.1 Gender differences in the propensity to collaborate in different forms, in the various disciplines*

Female and male academics belonging to the various UDAs show different propensities to collaborate. To analyze these differences, we present a table for each form of collaboration, showing per gender (F: female, M: male) and UDA: i) the average propensity to collaborate; ii) the percentage of academics with nil propensity; iii) percentage of academics with maximum (100 percent) propensity. The last column of Table 2 shows the results of the Mann-Whitney U test[8] (Mann and Whitney 1947), which is applied to verify the significance of the observed gender differences in collaboration. The initial analysis of the differences is conducted using the Wilcoxon test function (R Development Core Team, 2012); the sign + (-) in each cell highlights, for each UDA, if female academics have a higher (lower) median than males. The findings permit clustering of UDAs on the basis of the difference in propensities to collaborate of their female and male academics. Table 2 shows the values of propensity to collaborate, C. These generally appear extremely high both for female and male academics, in line with the results obtained by Abramo, D'Angelo, and Murgia (2013). In general, the propensity for women to collaborate is higher than that for their male colleagues. This result, further confirmed by the Mann-Whitney U test[9], is in agreement with what has been documented in previous research on scientific production by Italian academics over the 2001-2003 triennium (Abramo, D'Angelo, and Caprasecca 2008) and for production by Spanish academics in 2007 (Moya Anegón et al. 2009).

The large part of previous studies in the literature examine one or very few disciplines; however regardless of the breadth of the studies, they still show different results for the disciplines analyzed. For this reason, it is important to compare the propensity of male and female academics in a manner that takes account of their UDA specializations. As shown in Table 2, women and men register extremely high and relatively similar propensities to collaborate, with a slight difference in favor of women in all the UDAs (for Earth sciences and Civil engineering the differences are so slight as to not reach significance under the Mann-Whitney test). The results of our analysis thus agree with those of Kyvic and Teigen (1996), concerning scientific production in four Norwegian universities in the areas of Humanities and social sciences, Natural sciences, and Medical sciences, and with those of Long (1992), concerning publications resulting from mentor-fellow collaborations in Biochemistry. There is also partial agreement with the results from Prpic (2002), who analyzed the publications produced by a substantial sample of young Croat academics, showing that women seemed to have a greater propensity to collaborate than men, in the areas of Biotechnology and Social sciences and humanities. Still, Prpic's study, unlike our results, registers a lower propensity to collaborate for female versus male researchers in

---

[8] Although our dataset includes the entire population of Italian academics and is not a sample, we still apply the significance test for potential purposes of extending the results to other contexts and periods.
[9] The Mann-Whitney U test compares two samples, verifying the significance of the difference between the medians. For this reason there can be cases where the test shows a positive (or negative) difference between two samples even where the first sample has an average that is lower (higher) than the second (see the case of Civil engineering, Table 2).

Medicine, Natural sciences and Technical sciences. In the same fashion, our research results contrast with those reported by preceding studies concerning publications in the economics disciplines (Boschini and Sjogren 2007; McDowell and Smith 1992; McDowell et al. 2006), which showed a greater propensity to collaborate on the part of male researchers. Comparing the data from Table 2 and Table 1 we can see that in general the percentage of female researchers who collaborate is higher than for males, and that the percentage of their works in collaboration is consistently slightly higher. It is only for Physics, Mathematics and computer sciences, Pedagogy and psychology that the percentage of "collaborative" women is below that of men, but the women researchers still produce a higher percentage of co-authored publications than men.

*Table 2: Propensity to collaborate C, per UDA (percentage values)*

| UDA | Mean C | | % C = 0% | | % C = 100% | | U Mann-Whitney |
|---|---|---|---|---|---|---|---|
|  | F | M | F | M | F | M |  |
| Chemistry (CHE) | 99.6 | 98.9 | 0.0 | 0.2 | 96.7 | 93.7 | +*** |
| Medicine (MED) | 99.5 | 99.4 | 0.2 | 0.1 | 96.7 | 94.0 | +*** |
| Biology (BIO) | 99.4 | 98.8 | 0.1 | 0.1 | 96.7 | 92.1 | +*** |
| Agricultural and veterinary sciences (AVS) | 99.3 | 99.0 | 0.1 | 0.3 | 96.7 | 95.2 | +* |
| Physics (PHY) | 98.3 | 96.3 | 0.7 | 1.1 | 86.5 | 80.5 | +*** |
| Industrial and information engineering (IIE) | 98.1 | 96.9 | 0.3 | 0.5 | 89.8 | 84.8 | +*** |
| Earth sciences (EAR) | 97.8 | 97.6 | 0.7 | 0.7 | 91.8 | 90.2 | + |
| Pedagogy and psychology (PPS) | 96.8 | 96.6 | 2.1 | 0.6 | 92.5 | 86.7 | +** |
| Civil engineering (CEN) | 93.7 | 94.5 | 2.0 | 1.7 | 84.0 | 81.1 | + |
| Mathematics and computer sciences (MAT) | 90.6 | 88.4 | 3.0 | 3.4 | 74.7 | 65.7 | +*** |
| Economics and statistics (ECS) | 86.9 | 82.5 | 7.1 | 9.0 | 76.3 | 67.0 | +** |
| Total | 97.9 | 96.9 | 0.8 | 0.9 | 92.8 | 87.4 | +*** |

*Significance level: \*\*\*=p < 0.001; \*\*= p < 0.01; \*=p < 0.05*

The results for propensity to collaborate in general (*C*) do not discriminate potential differences concerning the different forms of collaboration. For this we add a deeper level of analysis of the three sub-forms of collaboration: intramural, extramural at the domestic level and extramural international. The results for intramural collaboration *CI*, are presented in Table 3. Both at the general level and the levels of the individual UDAs, the results show greater propensity for intramural collaboration on the part of women academics, except in Civil engineering where we identify greater propensity on the part of men, although not at a significant level. The result obtained in this UDA can be explained by the fact that, comparing genders, there are a greater percentage of female academics whose publications never result from intramural collaboration (11.0 percent vs. 8.3 percent), and above all there are a lower percentage of females whose publications are completely the result of intramural collaboration (45.0 percent vs. 46.5 percent). Apart from the data on Civil engineering, the greater propensity to intramural collaboration on the part of women could be interpreted as proof of the claim that, in the Italian university system, female academics are fully integrated within their universities. Clearly, to corroborate such a hypothesis it would be necessary to further verify the degree of gender homophily represented by such collaborations and above all the difference in academic rank between the women researchers and the male colleagues with whom they collaborate. The absence of studies in the literature on the different intramural collaborations of male and female colleagues do not permit comparison of the present results with those of other nations.

*Table 3: Propensity for intramural collaboration CI in the different UDAs (percentage values)*

| UDA | Mean CI | | % CI = 0% | | % CI = 100% | | U Mann-Whitney |
|---|---|---|---|---|---|---|---|
| | F | M | F | M | F | M | |
| Chemistry | 86.7 | 81.6 | 2.3 | 2.4 | 54.5 | 40.9 | +*** |
| Agricultural and veterinary sciences | 84.7 | 79.4 | 2.7 | 5.0 | 57.2 | 49.0 | +*** |
| Industrial and information engineering | 84.3 | 81.9 | 3.3 | 4.0 | 50.1 | 46.4 | +* |
| Medicine | 83.9 | 80.0 | 3.8 | 3.5 | 54.1 | 42.8 | +*** |
| Biology | 83.4 | 74.4 | 3.7 | 4.6 | 55.2 | 36.7 | +*** |
| Civil engineering | 70.2 | 74.0 | 11.0 | 8.3 | 45.0 | 46.5 | - |
| Physics | 74.9 | 65.0 | 6.8 | 9.1 | 35.8 | 27.8 | +*** |
| Earth sciences | 64.4 | 61.2 | 10.5 | 11.7 | 35.0 | 29.8 | + |
| Pedagogy and psychology | 62.2 | 56.6 | 20.5 | 15.5 | 41.3 | 29.4 | +* |
| Mathematics and computer sciences | 58.8 | 52.0 | 20.3 | 20.6 | 31.5 | 22.5 | +*** |
| Economics and statistics | 49.4 | 41.4 | 34.8 | 36.6 | 33.0 | 23.5 | +** |
| Total | 78.9 | 73.9 | 7.1 | 7.2 | 49.9 | 39.1 | +*** |

*Significance level: \*\*\*=p < 0.001; \*\*= p < 0.01; \*=p < 0.05*

Concerning domestic-level extramural collaboration there is at least one preceding study (Moya Anegón et al. 2009) that does permit a comparison with the results obtained from our study, as presented in Table 4. From this table we see that at the general level, women academics show a greater propensity to extramural domestic collaboration than do men.

However, descending to the UDA level, a number of highly differentiated situations emerge: Physics, Medicine, Earth sciences and Industrial and information engineering are the four disciplines where the greater propensity of women for domestic-level collaboration is confirmed, though only the last UDA has a gender difference that is statistically significant. The comparison of these results to those of Moya Anegón et al. (2009) does not provide unequivocal indications: these authors analyzed the differences between male and female academics in their collaboration with domestic (Spain) colleagues who are situated in different regions from the academic's home university. As in the current results, their study shows a higher propensity of women academics for domestic extramural collaborations in the two areas of Earth sciences and Medicine, but in Physics and Industrial and information it indicates a greater propensity for male colleagues. The result of greater propensity for males, in line with our findings, is also verified in the specialties related to Chemistry, Agricultural and veterinary sciences and Mathematics and computer sciences, while in the remaining UDAs, again contrary to our results, there is a greater propensity of women academics for domestic extramural collaboration.

*Table 4: Propensity for extramural collaboration at the domestic level CED in the different UDAs (percentage values)*

| UDA | Mean CED | | % CED = 0% | | % CED = 100% | | U Mann-Whitney |
|---|---|---|---|---|---|---|---|
| | F | M | F | M | F | M | |
| Physics | 73.0 | 72.5 | 4.6 | 5.8 | 25.8 | 24.7 | + |
| Medicine | 62.6 | 62.3 | 9.0 | 7.9 | 23.3 | 19.6 | + |
| Earth sciences | 60.4 | 58.0 | 12.6 | 13.3 | 25.5 | 22.5 | + |
| Biology | 57.1 | 57.7 | 11.4 | 8.2 | 19.1 | 15.9 | - |
| Chemistry | 49.2 | 50.1 | 7.9 | 8.5 | 9.9 | 9.0 | - |
| Agricultural and veterinary sciences | 46.9 | 47.1 | 15.3 | 18.3 | 12.5 | 14.5 | - |
| Pedagogy and psychology | 45.9 | 51.5 | 32.5 | 19.1 | 22.1 | 22.4 | -* |
| Economics and statistics | 38.5 | 37.8 | 41.3 | 37.7 | 21.2 | 18.1 | - |
| Mathematics and computer sciences | 33.6 | 33.6 | 38.4 | 30.9 | 13.0 | 9.6 | - |
| Civil engineering | 27.0 | 25.8 | 49.0 | 43.4 | 11.5 | 7.6 | - |
| Industrial and information engineering | 26.9 | 24.5 | 29.2 | 33.7 | 4.9 | 5.3 | +** |
| Total | 52.0 | 49.5 | 16.3 | 17.3 | 17.7 | 14.9 | +*** |

*Significance level: \*\*\*=p < 0.001; \*\*= p < 0.01; \*=p < 0.05*

The results concerning the propensity for international extramural collaboration are more homogeneous for the various UDAs, as shown in Table 5. In fact only Earth sciences registers a greater propensity for women, while at the general level and in all the other UDAs the propensity among men is greater, although often with Mann-Whitney scores that are not statistically significant. These results are in line with those recorded by Larivière et al. (2011), concerning scientific production of Canadian academics from 2000 to 2008, and by Frehill, Vlaicu, and Zippel (2010), from a survey of 103 U.S. professors responsible for international projects. Moya Anegón et al. (2009) also show a greater propensity for international collaboration on the part of male researchers in all specialties, except those related to Biology and Pedagogy and psychology, The lesser propensity to international collaboration on the part of women academics testifies that the barriers highlighted in the literature – prejudices in certain countries, difficulty in accessing funds, restrictions due to family commitments – continue to have a notable impact, in spite of the fact that in at least some countries there has been a rebalancing of family duties between the sexes.

The comparison of values of *C*, *CI*, *CED* and *CEF* for female and male academics, further articulated by the Mann-Whitney test, permits identification of differences and similarities between the UDAs. Earth sciences is the only UDA where women register a propensity higher than that of men for all the forms of collaboration analyzed. In Medicine, Physics and Industrial and information engineering, women register a propensity to collaborate higher than their male colleagues in all forms of collaboration analyzed, except international extramural. In Biology, Mathematics and computer sciences, Chemistry, Agricultural and veterinary sciences, Pedagogy and psychology, Economics and statistics, women register a higher propensity than men for collaboration at the general and intramural level, but lower at the extramural domestic and international levels. Finally, only in Civil Engineering, in spite of a propensity to collaborate that is generally higher than for male colleagues, female academics show a lower propensity for all of the sub-forms of collaboration.

*Table 5: Propensity to collaborate at the international extramural level CEF in the different UDAs (percentage values)*

| UDA | Mean CEF | | % CEF = 0% | | % CEF = 100% | | U Mann-Whitney |
|---|---|---|---|---|---|---|---|
| | F | M | F | M | F | M | |
| Physics | 48.7 | 52.3 | 14.6 | 11.7 | 11.3 | 9.7 | -* |
| Earth sciences | 35.6 | 31.9 | 31.0 | 35.4 | 10.5 | 7.0 | + |
| Pedagogy and psychology | 31.1 | 32.6 | 42.9 | 37.0 | 13.5 | 9.7 | - |
| Economics and statistics | 26.6 | 28.0 | 55.9 | 50.6 | 13.4 | 11.5 | - |
| Biology | 25.4 | 28.8 | 33.1 | 24.8 | 3.8 | 3.4 | -*** |
| Mathematics and computer sciences | 25.0 | 27.8 | 45.2 | 37.0 | 7.4 | 6.1 | -*** |
| Chemistry | 24.2 | 25.8 | 28.2 | 25.4 | 0.9 | 1.5 | -* |
| Medicine | 19.1 | 18.4 | 44.0 | 40.8 | 3.7 | 2.6 | - |
| Agricultural and veterinary sciences | 18.5 | 20.9 | 45.0 | 43.2 | 2.5 | 3.5 | -* |
| Civil engineering | 14.6 | 15.4 | 62.0 | 58.8 | 4.0 | 2.6 | - |
| Industrial and information engineering | 13.3 | 13.4 | 50.7 | 49.9 | 1.8 | 1.5 | - |
| Total | 23.6 | 23.9 | 39.3 | 37.5 | 4.7 | 3.9 | - |

*Significance level: \*\*\*=p < 0.001; \*\*= p < 0.01; \*=p < 0.05*

The differences between men and women academics in their forms of collaboration thus varies according to UDA. This could be due to certain factors that characterize each discipline, beginning from the percentage of women in the total research staff. In this case, in the presence of gender homophily we would expect a greater propensity for women to collaborate in the UDAs where their presence is more substantial. However in our analysis, for the UDAs with the highest percentages of female academics, the greater propensity to collaborate with respect to male colleagues emerges only at the intramural level.

*4.2 Gender differences in the correlation between propensities to collaborate in different forms*

Development of a publication can require initiation of collaborations that are both intramural and extramural, for example when it is necessary to share resources with colleagues in the same university, and at the same time gather data on samples from further contexts. In other cases a researcher could be motivated to limit collaboration to a single form, for example because he or she would not want to share "ownership" of publications with too many people, or to avoid excessive costs for coordination. As demonstrated by Abramo, D'Angelo, and Murgia (2013), the disciplinary specialty of the researcher has a substantial effect on the choice to initiate one or more forms of collaboration. In this work we are interested in determining if there are relevant differences between the men and women researchers of each UDA in regards to these choices. To continue the analysis we now apply the *R rcorr function* to calculate the Spearman non-parametric correlation between the values obtained from each academic for the four indicators *C*, *CI*, *CED* and *CEF*. The results, presented in Table 6, permit comparison of female and male academics, both at the general level and for each UDA, showing if and how the four forms of collaboration are correlated among each other.

The analysis presented in Table 6 shows how the correlation values, both for female and male academics, are generally strongly significant, although often below 0.40. In particular, the propensity to collaborate *C* results as positively and strongly correlated (Spearman correlations near 0.4) to *CI* in only certain UDAs (Civil Engineering, Economy and statistics, Mathematics and computer sciences), with the correlation for male academics also slightly higher than for females. The correlation between *CI* and *CED* is instead seen to be negative, both for females and males; the correlation reaches strong levels in only two UDAs (Civil engineering, Industrial and information engineering), where female academics register slightly stronger correlations. In contrast, the correlation between *CI* and *CEF* never reaches strong levels, although it is always negative, both for men and women. Finally, the correlation between *CED* and *CEF* assumes positive and strong values only in Physics, a UDA where female academics also register a higher correlation than male colleagues.

*Table 6: Spearman correlation between the indicators of propensity to collaborate, per UDA*

| UDA | Gender | C-CI | C-CED | C-CEF | CI-CED | CI-CEF | CED-CEF |
|---|---|---|---|---|---|---|---|
| AVS | F | 0.22*** | 0.11*** | 0.03 | -0.32*** | -0.23*** | 0.01 |
|  | M | 0.20*** | 0.10*** | 0.06** | -0.36*** | -0.26*** | -0.01 |
| BIO | F | 0.19*** | 0.11*** | 0.03 | -0.24*** | -0.19*** | 0.01 |
|  | M | 0.24*** | 0.14*** | 0.02 | -0.19*** | -0.30*** | -0.07*** |
| CEN | F | 0.41*** | 0.16* | 0.13 | -0.45*** | -0.23** | -0.04 |
|  | M | 0.44*** | 0.14*** | 0.10** | -0.44*** | -0.25*** | 0.06 |
| CHE | F | 0.17*** | 0.09** | 0.02 | -0.31*** | -0.18*** | -0.02 |
|  | M | 0.27*** | 0.14*** | 0.06** | -0.26*** | -0.23*** | 0.00 |
| EAR | F | 0.22*** | 0.19** | 0.10 | -0.17** | -0.27*** | 0.04 |
|  | M | 0.21*** | 0.21*** | 0.11** | -0.32*** | -0.25*** | 0.09** |
| ECS | F | 0.39*** | 0.31*** | 0.25*** | -0.28*** | -0.24*** | 0.02 |
|  | M | 0.42*** | 0.37*** | 0.29*** | -0.16*** | -0.22*** | -0.07 |
| IIE | F | 0.34*** | 0.09** | 0.05 | -0.46*** | -0.28*** | 0.17*** |
|  | M | 0.39*** | 0.12*** | 0.06*** | -0.43*** | -0.32*** | 0.15*** |
| MAT | F | 0.40*** | 0.24*** | 0.19*** | -0.37*** | -0.26*** | 0.00 |
|  | M | 0.44*** | 0.25*** | 0.21*** | -0.29*** | -0.32*** | 0.05* |
| MED | F | 0.18*** | 0.11*** | 0.02 | -0.25*** | -0.24*** | 0.07*** |
|  | M | 0.16*** | 0.09*** | 0.01 | -0.30*** | -0.28*** | 0.07*** |
| PHY | F | 0.25*** | 0.23*** | 0.11* | -0.14** | -0.04 | 0.40*** |
|  | M | 0.33*** | 0.36*** | 0.22*** | 0.01 | -0.03 | 0.36*** |
| PPS | F | 0.24*** | 0.2*** | 0.15** | -0.36*** | -0.27*** | 0.02 |
|  | M | 0.23*** | 0.23*** | 0.12* | -0.21*** | -0.34*** | -0.10 |
| Total | F | 0.33*** | 0.20*** | 0.08*** | -0.20*** | -0.23*** | 0.07*** |
|  | M | 0.36*** | 0.21*** | 0.08*** | -0.22*** | -0.29*** | 0.15*** |

*Significance level: \*\*\*=p < 0.001; \*\*= p < 0.01; \*=p < 0.05*

## 5. Conclusions

The scientific debate over the role of women in the academic system has brought attention to a set of factors that seem to contribute to broad problems still existing in many countries: from low percentages of women academics to discrimination that affects career opportunities, to a "productivity gap" compared to men. Thanks in part to the analyses reported in the literature, policies have been formulated and specific interventions adopted to act at the level of the underlying factors and reduce gender differences in research systems.

In spite of the increasing reliance on collaboration in research activity, studies on differences in the collaborations achieved by female and male academics are still focused on limited contexts, and above all have not investigated all the possible forms of collaboration. The lack of more thorough inquiry seems surprising, since some of the limiting factors for women academics, such as various prejudices, difficulties in accessing funds and limitations caused by family responsibilities, clearly have different impacts on the potential forms in which women can develop their collaborations, and thus their scientific productivity and careers.

The present study attempts to respond to the gap in research, applying an approach that takes the individual scientist as the base analytical unit, and thus permits accurate measurement of the propensity of individual academics to initiate each form of collaboration. The approach provides a realistic perception of the behavior of researchers in the overall system and disciplines analyzed, as well as comparisons between female and male academics that are not distorted by outliers, i.e. by single scientists who may alone develop high numbers of collaborations and thus heavily influence the values of aggregate indexes.

The application of our methodology to the scientific production of Italian academics demonstrates that female researchers, in contrast to what has been indicated in some of the previous literature, show greater propensity to collaborate at the general level, intramural level and domestic extramural level. However a gap remains in the propensity to collaborate at the international level, suggesting the need to reinforce mitigative policies, such as those permitting women academics, and their families, to increase their mobility beyond national confines. The findings also demonstrate the need for caution in administering evaluations that draw on indicators of international collaboration, which would clearly penalize the female gender and organizations with higher concentrations of women academics.

The analysis further shows how the gap for women academics, as well as the correlation among the propensities for various forms of collaboration, is linked to their discipline of specialization. This is undoubtedly related to the emphasis on various forms of collaboration typical of each discipline, with their follow-on effects for female academics, but could also be a legacy of academic cultures that are generally less open to collaboration with women. To verify this hypothesis it would be useful to analyze the structure of the collaborations actuated by female and male academics, in order to detect whether or not there are mechanisms of gender homophily and whether or not collaborations are characterized by differences in rank between academics. The authors will examine these more detailed questions in further research.